\documentclass[12pt]{article}
\usepackage[margin=1in]{geometry}
\usepackage[small]{titlesec}
\usepackage{natbib,amsmath,amssymb,amsthm,graphicx,setspace,paralist,booktabs,rotating,
	subcaption,float}
\usepackage{times}
\usepackage{multirow}

\AtBeginDocument{%
	\abovedisplayskip=5pt plus 2pt minus 2pt
	\belowdisplayskip=\abovedisplayskip
	\abovedisplayshortskip=2pt plus 2pt minus 2pt
	\belowdisplayshortskip=\belowdisplayskip}
\titlespacing*{\section}{0pt}{*2}{*1}
\titlespacing*{\subsection}{0pt}{*2}{*1} 
\bibpunct{(}{)}{;}{a}{}{,}
\usepackage{threeparttable}
\usepackage{hyperref}
\usepackage[usenames]{color}
\usepackage{enumerate}

\def\argmin{\mathop{\rm arg\, min}}
\def\R{\mathbb{R}}
\def\E{\mathbb{E}}
\def\P{\mathbb{P}}

\def\eps{\epsilon}
\def\Sig{\Sigma}
\def\sig{\sigma}

\def\lam{\lambda}
\def\gam{\gamma}

\newtheorem{condition}{Condition}[section]
\newtheorem{theorem}{Theorem}[section]

\newtheorem{lemma}{Lemma}[section]
\newtheorem{proposition}{Proposition}[section]
\newtheorem{remark}{Remark}[section]

\usepackage{bm}
\newcommand{\ignore}[1]{}
 
\def\gam{\gamma}
\def\bgam{\bm{\gam}}
\def\beps{\bm{\eps}}
\def\bX{\mathbf{X}}
\def\bZ{\mathbf{Z}}
\def\by{\mathbf{y}}

\def\bSig{\mathbf{\Sigma}}

\allowdisplaybreaks[1]

\begin{document}
	

	\begin{titlepage}
	\setstretch{1.24}
	\title{Inference for high-dimensional linear mixed-effects models: A quasi-likelihood approach}
	\author{Sai Li, \ T. Tony Cai \ and \ Hongzhe Li}
	\date{}
	\maketitle
	\thispagestyle{empty}
	\footnotetext{Sai Li is a postdoctoral researcher at Department of Biostatistics, Epidemiology and Informatics, Perelman School of Medicine, University of Pennsylvania, Philadelphia, PA 19104 (E-mail: \emph{Sai.Li@pennmedicine.upenn.edu}). T. Tony Cai is Daniel H. Silberberg Professor of Statistics, Department of Statistics, The Wharton School, University of Pennsylvania, Philadelphia, PA 19104 (E-mail:\emph{tcai@wharton.upenn.edu}). Hongzhe Li is Perelman Professor of Biostatistics, Department of Biostatistics, Epidemiology and Informatics, Perelman School of Medicine, University of Pennsylvania, Philadelphia, PA 19104 (E-mail: \emph{hongzhe@upenn.edu}).}

\begin{abstract}
Linear mixed-effects models are widely used in analyzing clustered or repeated measures data. We propose a quasi-likelihood approach for estimation and inference of the unknown parameters in linear mixed-effects models with high-dimensional fixed effects. The proposed method is applicable to general settings where the dimension of the random effects and the cluster sizes are possibly large. Regarding the fixed effects, we provide rate optimal estimators and valid inference procedures that  do not rely on the structural information of the variance components. We also study the estimation of  variance components with high-dimensional fixed effects in general settings. The algorithms are easy to implement and computationally fast. The proposed methods are assessed in various simulation settings and are applied to a real study regarding the associations between  body mass index and genetic  polymorphic markers in a heterogeneous stock mice population.

	\bigskip

\noindent%
	\end{abstract}
  
\end{titlepage}
\section{Introduction}
The results of scientific experiments are often subject to environmental effects as  experimental units can be  grouped and settled in diverse environments, where the observations within the same group can be dependent as a cluster.
Clustered data commonly arise in many fields, such as biology, genetics, and economics. Linear mixed-effects models provide a flexible tool for analyzing such clustered data, which include repeated measures data, longitudinal data, and multilevel data \citep{PB00, Goldstein93}. 
The linear mixed-effects models incorporate both the fixed and random effects, where the random effects induce correlations among the observations within each cluster and accommodate the cluster structure. In many genomic and economic studies, the dimension of the covariates can be large and possibly much larger than the sample size. A variety of statistical models and approaches have been proposed and studied for analyzing high-dimensional data. However, most of them are restricted to dealing with independent observations, such as linear models and generalized linear models. Statistical inference for high-dimensional linear mixed-effects models remains to be a challenging problem. In this work, we consider estimation and inference of unknown parameters in high-dimensional mixed-effects models.

For ease of presentation, we use the setting for clustered data to present a  linear mixed-effects model.  For repeated measurement data, the repeated measures form a cluster. Let $i=1,\dots,n$ be the cluster indices. For the $i$-th cluster, we have a  response vector $y_i\in\R^{m_i}$, a design matrix for the fixed effects $X^i\in\R^{m_i\times p}$, and a design matrix for the random effects $Z^i\in\R^{m_i\times q}$, where $m_i$ is the size of the $i$-th cluster. A linear mixed-effects model \citep{LW82} can be written as 
\begin{align}
\label{m1}
y_i=X^i\beta^*+Z^i\gamma_i+\eps_i,~i=1,\dots, n,
\end{align}
where $\beta^*\in\R^p$ is the vector of the fixed effects, $\gam_i\in\R^q$ is the vector of the random effects of the $i$-th cluster, and $\eps_i\in\R^{m_i}$ is the noise vector of the $i$-th cluster. For $i=1,\dots,n$, we assume $\gam_i$ and $\eps_i$ are independently distributed with mean zero and variance $\Psi\in\R^{q\times q}$ and $\sig^{2}_eI_{m_i}$, respectively. Detailed assumptions are given in Sections \ref{sec2} and \ref{sec3}.

Much existing literature on linear mixed-effects models assumes that the number of random effects $q$ and cluster sizes $m_i$ are fixed. Without special emphasis, we say a fixed-dimensional setting if $p$, $q$, and $\{m_i\}_{i=1}^n$ are all fixed numbers, and a high-dimensional setting if $p$ is large and possibly much larger than $N$, where $N=\sum_{i=1}^n m_i$ is the total sample size. We refer to $\gam_i$ and $\eps_i$ as the random components.

\subsection{Related literature}
\label{sec1-1}
In the fixed-dimensional setting, many methods have been proposed to jointly estimate the fixed effects and variance parameters. We refer to \citet{GD11} for a comprehensive review. Among them, the maximum likelihood estimators (MLEs) and restricted MLEs are most popular for estimation and inference in linear mixed-effects models. Restricted MLEs can produce unbiased estimators of the variance components in the low-dimensional setting but it is not applicable in the high-dimensional setting.  
Furthermore, these likelihood-based estimators rely heavily on the normality assumptions of the random components. Computationally, maximizing the likelihood can generally lead to a nonconvex optimization problem that  typically has multiple local maxima. Hence, the performance of likelihood-based methods  lacks of guarantees in real applications. 

As an alternative, \citet{SZT07} proposed moment estimators of the fixed effects and variance parameters for a random effect varying-coefficient model.  \citet{PY12} considered such moment estimators for fixed-dimensional linear mixed-effects models. Their proposed estimators have closed-form solutions and are computationally efficient. The consistency and asymptotic normality of these estimators are justified under certain conditions in the fixed-dimensional setting. 
\citet{Ahmn12} proposed another moment-based method for the estimation and selection of the variance components of the random effects in the fixed-dimensional setting. This method works especially well when the number of the random effects is as large as the cluster sizes, i.e. $m_1=\dots=m_n=q$.
 
 For inference of variance components in the fixed-dimensional setting, the likelihood ratio, score, and Wald tests \citep{SL94, Lin97, VM03, D04} are broadly used. However, when testing the existence of the random effects, the asymptotic distribution of the likelihood ratio is usually a mixture of chi-square distributions \citep{Miller77, Self87}. Since these methods are based on the MLEs or restricted MLEs as initial estimators, they also suffer from the drawbacks of likelihood-based methods discussed above.  

In the high-dimensional setting, the problems are much more challenging. Assuming fixed cluster sizes, \citet{SBV10} analyzed the rate of convergence for the global maximizer of the $\ell_1$-penalized likelihood with fixed designs. As mentioned before, the analysis for the global optimum may not apply to the realizations due to the existence of local maxima. \citet{FanLi12} studied the fixed effects and random effects selection in a high-dimensional linear mixed-effects model when the cluster sizes are balanced, i.e. $(\max_i m_i)/(\min_i m_i)<\infty$. The selection consistency requires minimum signal strength conditions regarding the fixed effects and the random effects. \citet{Bradic17} considered testing a single coefficient of the fixed effects in the high-dimensional linear mixed-effects models with fixed cluster sizes, fixed number of random effects, and sub-Gaussian designs. The theoretical analyses in all three aforementioned papers require the positive definiteness of the covariance matrix of the random effects. This condition takes prior knowledge on the existence of the random effects and can be hard to fulfill in applications. Moreover, the optimal convergence rate of parameter estimation remains unknown. In fact, estimators of fixed effects in \citet{SBV10} and \citet{Bradic17} may not be rate optimal for estimation according to our analysis. Finally,  estimation and inference  of the variance components in the high-dimensional setting remain  largely unknown.

The problems of estimation and inference of the fixed effects in linear mixed-effects models are related to  high-dimensional linear models. Many penalized methods have been proposed for prediction, estimation, and variable selection in high-dimensional linear models; see, for example, \citet{Lasso, FL01, Zou06, CT07, MB10, Zhang10}. Statistical inference on a low-dimensional component of a high-dimensional regression vector has been considered and studied in linear models and generalized linear models with ``debiased'' estimators \citep{ZZ14, van14, JM14}, and the minimaxity and adaptivity of confidence intervals have been studied in \citet{CG17} and \citet{Cai2020CI-GLM}.
The idea of debiasing has also been studied and extended to solve other statistical problems, such as statistical inference in Cox models \citep{Fang17}, simultaneous inference \citep{Zhang17, Dezeure17}, and semi-supervised inference \citep{CaiGuo2020CHIVE}. 

\subsection{Our contributions} 
In this paper, we develop a simple but powerful method for inference of the unknown parameters in high-dimensional linear mixed-effects models. 
Our method  is applicable to the settings where the number of random effects can possibly  be large and the cluster sizes can be either fixed or growing, balanced or unbalanced. The proposed method is easy to implement and the optimization in each step is either analytic or convex.

Based on a proxy of the true covariance matrix, we develop a penalized quasi-likelihood approach for fixed effects estimation. The proposed estimator is minimax rate optimal under general conditions. We further develop a debiased estimator for hypothesis testing and construction of confidence intervals for the fixed effects. The proposed estimator does not require  normality assumptions or the structural assumptions on the variance components. 
We further apply the idea of quasi-likelihood to estimate the variance components and prove its optimality under certain conditions.

 Our analysis provides a novel insight  for understanding and simplifying the linear mixed-effects models by approximating the true unknown covariance matrix of the random components with some simple proxy matrices. In this way, one separates the tasks of estimating the fixed effects and variance components and avoids  the nuisance parameters in each optimization step. This improves the computational efficiency and simplifies the theoretical analysis. 

\subsection{Notation} Throughout the paper, we use $i$ to index the $i$-th cluster and $k$ to index the $k$-th observation in each cluster. Let $\by$, $\bgam$, $\beps$, and $\bX$ be obtained by stacking vectors $y_i$, $\gam_i$, $\eps_i$, and matrices $X^i$ underneath each other, respectively. Let $\bZ\in\R^{N\times (nq)}$ be a block diagonal matrix with the $i$-th block being $Z^i$. Let $\Sig^i_{\theta}=Z^i\Psi Z^i+\sig^2_eI_{m_i}$ and $\bSig_{\theta}\in\R^{N\times N}$ be a block diagonal matrix with the $i$-th block being $\Sig^i_{\theta}$. Let $\Sig^i_{z}=(Z^i)^{\top}Z^i/m_i$ and $\Sig^i_{z,x}=(Z^i)^{\top}X^i/m_i$, $i=1,\dots,n$.
For a random variable $u\in\R$, define its sub-Gaussian norm as $\|u\|_{\psi_2}=\sup_{l\geq 1} l^{-1/2}\E^{1/l}[|u|^l]$. We refer to $\|u\|_{\psi_2,\bZ}=\sup_{l\geq 1} l^{-1/2}\E^{1/l}[|u|^l|\bZ]$ as the conditional sub-Gaussian norm of $u$. 
For a random vector $U\in\R^{n_0}$, define its sub-Gaussian norm as $
   \|U\|_{\psi_2}=\sup_{\|v\|_2=1,v\in\R^{n_0}}\|\langle U,v\rangle\|_{\psi_2}$. Define it conditional sub-Gaussian norm as $\|U\|_{\psi_2,\bZ}=\sup_{\|v\|_2=1,v\in\R^{n_0}}\|\langle U,v\rangle\|_{\psi_2,\bZ}$.

Let $A\in\R^{n_0\times n_0}$ be a symmetric matrix. $A\succeq 0$ means that $A$ is semi-positive definite and $A\succ 0$ means that $A$ is positive definite. Let $\Lambda_{\max}(A)$ and $\Lambda_{\min}(A)$ denote the largest and smallest eigenvalues of $A$, respectively. Let $\|A\|_2$ denote $\Lambda_{\max}(A)$. Let $\|A\|_F^2=\textrm{Tr}(A^{\top}A)$, where $\textrm{Tr}(A)$ is the trace of matrix $A$.  
Let $c,c_0,c_1,\dots, C, C_0,C_1,\dots$ denote some generic positive constants that  can vary in different statements.

\subsection{Organization of the paper}
The rest of the paper is organized as follows.  Section \ref{sec2} introduces the idea of quasi-likelihood and a procedure for the fixed effects inference. Section \ref{sec3}  provides a theoretical analysis for the inference procedures proposed in Section \ref{sec2}. Section \ref{sec4} introduces estimators for the variance components and provides upper and lower bounds. Numerical performance of the proposed methods is investigated in Section \ref{sec5} in various simulation settings. The proposed methods are applied in Section \ref{sec6} to analyze a real study on the associations between  the body mass index and genetic variants  in a stock mice population where the cage effect is modeled as a random effect.  A discussion is given in Section \ref{sec-diss}. Proofs and more numerical results are provided in the Supplementary Materials.

\section{Inference for fixed effects: the method}
\label{sec2}
In many applications of the linear mixed-effects models, inference of the fixed effects is of main interest.
In this section, we present our method  for fixed effects inference and describe its motivations.  We assume that the vector of fixed effects $\beta^*$ is sparse such that $\|\beta^*\|_0\leq s$ with $s$ unknown. We consider model (\ref{m1}) where $p$, $s$, and $q$ can grow and $p$ can be much larger than $N$. The cluster sizes $\{m_i\}_{i=1}^n$ can be either fixed or grow with $n$. 

\subsection{Motivations of the proposed method}
For fixed effects estimation in model (\ref{m1}), the main challenges are posed by the high-dimensionality of the fixed effects and the clustered structure of the observations. Before developing a new method, it is helpful to understand the new challenges posed by the cluster structures in model (\ref{m1}) in terms of estimation and inference. For this purpose, we study the consequences of mis-specifying a linear mixed-effects model as  a standard linear model. 

Applying Lasso \citep{Lasso} directly to the observations generated from (\ref{m1}), we analyze
\begin{equation}
\label{beta-naive}
\hat{\beta}^{(lm)}=\argmin_{b\in\R^p}\left\{\frac{1}{2N}\|\by-\bX b\|_2^2+ \lambda^{(lm)}\|b\|_1\right\}
\end{equation}
for some tuning parameter $\lambda^{(lm)}>0$. In a typical analysis of the Lasso, the convergence rate of $\hat{\beta}^{(lm)}$ depends on the restricted isometries of the sample covariance matrix, $\bX^{\top}\bX/N$, the sparsity of the true coefficients, and the so-called ``empirical process'' part of the problem, $\|\bX^{\top}(\by-\bX\beta^*)/N\|_{\infty}$. It is known that for linear models with row-wise independent sub-Gaussian $(\bX,\by)$, the ``empirical process'' part is of order $\sqrt{\log p/N}$, which gives the optimal convergence rate in $\ell_2$-norm. In the following proposition, we study the size of ``empirical process'' part when the true model is (\ref{m1}). 

\begin{proposition}[The rate of  Lasso for linear mixed-effects models]	
\label{prop1}
$\quad$
Suppose that the responses $y_i$ are generated with respect to model (\ref{m1}) and each row of $\bX$ is independently generated with covariance matrix $\Sig_{x|z}$ conditioning on $\bZ$. Then for any fixed $j\in\{1,\dots,p\}$,
{\small
	\begin{align}
	\label{prop1-eq1}
	\E\left[\left|\frac{1}{N}\bX_{.,j}^{\top}(\bZ\bgam+\beps)\right|^2\big|\bZ\right]&= \frac{(\Sig_{x|z})_{j,j}\sig^2_e}{N}+\frac{(\Sig_{x|z})_{j,j}\sum_{i=1}^nm_i\textup{Tr}(\Psi\Sig^i_{z})}{N^2} + \frac{\sum_{i=1}^nm_i^2 \|\Psi^{1/2}\E[\Sig^i_{z,x}|\bZ] \|_2^2}{N^2}.
	\end{align}
	}
\end{proposition} 
If $\Psi$ is positive definite and $\{\Sig^i_{z}\}_{i=1}^n$ have bounded diagonal elements, then the second term on the right hand side of (\ref{prop1-eq1}) is $\asymp q/N$. If it further holds that $\E[\Sig^i_{z,x}|\bZ]\neq 0$, i.e. $\bX$ and $\bZ$ are correlated, then the third term can be $\gtrsim \min_{1\leq i\leq n}m_i/N$. 
That is, the Lasso may not be rate optimal for clustered data if either $q$ grows, or, $\{m_i\}_{i=1}^n$ grow and $\bX$ and $\bZ$ are correlated. On the other hand, if the $q$ and $m_i$'s are all constant, it is not hard to prove that the original Lasso is still rate optimal for model (\ref{m1}). 

Therefore, proper methods need to be developed for high-dimensional linear mixed-effects models under general conditions on $q$ and $\{m_i\}_{i=1}^n$.  The main challenge comes from the correlation among observations induced by the random effects. For the $i$-th block, the covariance of the random components is $\Sig^i_{\theta^*}=Z^i\Psi(Z^i)^{\top}+\sig^2_eI_{m_i}$, which involves unknown parameters.  We consider a proxy of $\Sig^i_{\theta^*}$ as
\[
\Sig^i_a=aZ^i(Z^i)^{\top}+I_{m_i}
\]
with some predetermined constant $a> 0$. 
The following proposition shows that this approximation is valid up to some scaling constant. Let $\bSig_a\in\R^{N\times N}$ be the block diagonal matrix with the $i$-th block being $\Sig^i_a$.
\begin{proposition}
	\label{prop2}
	If $\Psi$ is positive definite, then for any constant $a>0$,
	\[
	\min\left\{\frac{1}{\sig^2_e}, \frac{a}{\Lambda_{\max}(\Psi)}\right\}\bSig^{-1}_{a}\preceq \bSig^{-1}_{\theta^*}\preceq  \max\left\{\frac{1}{\sig^2_e}, \frac{a}{\Lambda_{\min}(\Psi)}\right\}\bSig^{-1}_{a}.
	\]
\end{proposition}
Therefore, if $\Psi$ has positive and bounded eigenvalues, $\bSig^{-1}_{\theta^*}$ and $\bSig^{-1}_a$ are of the same rate and only differ by constants. This property of $\bSig^{-1}_a$ is crucial to achieve the general results in this work. A broader class of proxy matrices have been considered in \citet{FanLi12} for variable selection and in \citet{Bradic17} for hypothesis testing, which include $\bSig^{-1}_a$ as a special case. As reviewed in Section \ref{sec1-1},  afore-mentioned two papers considered relatively restrictive scenarios in terms of group sizes and the dimension of the random effects. It is not clear whether the desired property proved in Proposition \ref{prop2} holds for the general class of proxy matrices.

  

\subsection{The quasi-likelihood approach}
\label{sec2-1}
We consider a quasi-likelihood approach which replaces $\Sig^i_{\theta^*}$ with $\Sig^i_a$ for some constant $a>0$ in the likelihood function for Gaussian mixed-effects models. Specifically, let $\bX_a$ and $\by_a$ denote the transformed observations such that $(\bX_a,\by_a)=(\bSig^{-1/2}_a\bX,\bSig^{-1/2}_a\by)$.  

First, we estimate the fixed effects via the Lasso based on the transformed data. For some fixed $a>0$, define
\begin{equation}
\label{beta-mc}
   \hat{\beta}=\argmin_{\beta\in\R^p} \left\{\frac{1}{2\textup{Tr}(\bSig^{-1}_a)} \|\by_a-\bX_a\beta\|_2^2+\lam\|\beta\|_1\right\}
   \end{equation}
for some tuning parameter $\lam>0$. 
The quantity $\textup{Tr}(\bSig^{-1}_a)$ can be viewed as the effective sample size in the current problem and its magnitude is studied in Remark \ref{re3-1}. 
The choice of $a$ will be studied theoretically in Section \ref{sec3} and numerically in Section \ref{sec5}. 

Given the task of making inference for $\beta^*_j$, we propose the following debiased estimator. For $\hat{\beta}$ defined in (\ref{beta-mc}),
\begin{equation}
\label{eq-db}
\hat{\beta}^{(db)}_j= \hat{\beta}_j+ \frac{\hat{w}_j^{\top}(\by_a-\bX_a\hat{\beta})}{\hat{w}_j^{\top}(\bX_a)_{.,j}},
\end{equation}
where $\hat{w}_j\in\R^N$ can be viewed as a correction score. It can be computed via another Lasso regression \citep{ZZ14, van14} or via a quadratic optimization \citep{ZZ14, JM14}. For computational convenience, we consider the Lasso approach based on the transformed data. Define the correction score $\hat{w}_j=(\bX_a)_{.,j}-(\bX_a)_{.,-j}\hat{\kappa}_{j}$, where
\begin{align}
\label{eq-kappa}
    \hat{\kappa}_{j}=\argmin_{\kappa_{j}\in\R^{p-1}} \left\{\frac{1}{2\textup{Tr}(\bSig^{-1}_a)}\left\|(\bX_a)_{.,j}-(\bX_a)_{.,-j}\kappa_j\right\|_2^2 + \lam_j \|\kappa_{j}\|_1\right\},
\end{align}
for some tuning parameter $\lam_j>0$. A two-sided $100\times (1-\alpha)\%$ confidence interval for $\beta_j^*$ can be constructed as
\begin{equation}
\label{ci1}
   \hat{\beta}^{(db)}_j\pm z_{\alpha/2}\sqrt{\widehat{V}_j},
\end{equation}
where $z_\tau$ is the $\tau$-th quantile of a standard normal distribution and $\widehat{V}_j$ is an estimator of the variance of $\hat{\beta}^{(db)}_j$. We propose to use the following empirical variance estimate
\begin{equation}
\label{eq-var-db}
\widehat{V}_j= \frac{\sum_{i=1}^n\left[(\hat{w}^i_j)^{\top}(y_a^i-X_a^i\hat{\beta})\right]^2}{(\hat{w}_j^{\top}(\bX_a)_{.,j})^2},
\end{equation}
where $\hat{\beta}$ is the initial Lasso estimator (\ref{beta-mc}), $\hat{w}^i_j\in\R^{m_i}$ is the $i$-th sub-vector of $\hat{w}_j$ such that $\hat{w}_j=((\hat{w}^1_j)^{\top},\dots,(\hat{w}^n_j)^{\top})^{\top}$, and $y^i_a$ is the $i$-th sub-vector of $\by_a$. The idea of empirical variance estimator has been considered in \citet{Bu15} to deal with the misspecified linear models. The format of (\ref{eq-var-db}) is however different from the one for linear models because it is an average over $n$ groups rather than $N$ observations. In this work, $\widehat{V}_j$ serves as a convenient alternative to the limiting distribution-based variance estimators. In fact, the limiting distribution of $\hat{\beta}^{(db)}_j$ involves nuisance parameters coming from the complicated variance components. By using the empirical residuals of the transformed data, we bypass the estimation of the nuisance parameters. 

\section{Inference for fixed effects: theoretical guarantees}
  \label{sec3}
In this section, we provide theoretical guarantees for the estimators described in Section \ref{sec2-1}.  We first detail our assumptions.
\begin{condition}[Sub-Gaussian random components]
\label{cond1}
The random noises $\eps_{i,k},~i=1,\dots,n$, $k=1,\dots,m_i$, are independent with mean zero and variance $0<\sig^{2}_e<K_0<\infty$. The sub-Gaussian norms of $\eps_{i,k}$ are upper bounded by $K_0$.
The random effects $\gam_i\in\R^q,~i=1,\dots,n$, are independent with mean zero and covariance $\Psi\preceq K_1I_q$ for some positive constant $K_1$. For $i=1,\dots,n$, $\eps_i$ and $\gam_i$ are independent of each other and are independent of $(X^i,Z^i)$. The sub-Gaussian norms of $\bSig_{\theta^*}^{-1/2}(\bZ\bgam+\beps)$ are  upper bounded by $K_0$. \end{condition}
 Condition \ref{cond1} assumes sub-Gaussian random components while that classical linear mixed-effects models always assume Gaussian random components \citep{PB00}. Hence, Condition \ref{cond1} is less restrictive and is more robust to model misspecifications than the classical assumptions. In addition,  we do not require $\Psi$ to be strictly positive definite. A scenario of singular $\Psi$ is that some components of the random effects do not exists such that some diagonal elements of $\Psi$ are zero.

Regarding the conditions on the designs, the estimation and inference in the linear mixed-effects models are usually conditioning on $\bZ$ in order to maintain the cluster structure. \citet{SBV10} and \citet{FanLi12} assume both $\bX$ and $\bZ$ are fixed. \citet{Jiang16} considers estimation and inference in a misspecified linear model when both $\bX$ and $\bZ$ are random. \citet{Bradic17} assumes $\bX$ is sub-Gaussian with mean zero and $\bZ$ is fixed, which implies that $\bX$ and $\bZ$ are independent. In the current work, we consider random designs satisfying the following condition. 
\begin{condition}[Sub-Gaussian $\bX$ conditioning on $\bZ$]
\label{cond0a}
Conditioning on $\bZ$, each row of $\bX$ is independent with mean zero and covariance matrix $\Sig_{x|z}$ such that $0<K_*\leq \Lambda_{\min}(\Sig_{x|z})\leq \Lambda_{\max}(\Sig_{x|z})\leq K^*<\infty$. Conditioning on $\bZ$, the conditional sub-Gaussian norms of $X^i_{k,.}$ are upper bounded by $K_0$.
\end{condition}
In Condition \ref{cond0a},  we assume sub-Gaussian $\bX$ and $\bZ$  have mean independence, i.e., $\E[\bX|\bZ]=0$ for simplicity. This is slightly weaker than assuming $\bX$ and $\bZ$ are mutually independent and it holds when $\bZ$ is deterministic including the random intercept model.
In Section \ref{sec3-4}, we study the performance of our proposal when  $\E[\bX|\bZ]\neq 0$.

  \subsection{Fixed effects estimation}
In this subsection, we analyze the theoretical performance of (\ref{beta-mc}) under Conditions \ref{cond1} and \ref{cond0a}. Define 
\[
   \lam^*_{a}=\frac{\sqrt{\textup{Tr}(\bSig^{-1}_a\bSig_{\theta^*}\bSig^{-1}_a)\log p}}{\textup{Tr}(\bSig^{-1}_a)}.
\]
\begin{lemma}[Fixed effects estimation with quasi-likelihood based Lasso]
\label{lem1a}
Assume that Conditions \ref{cond1} and \ref{cond0a} hold true. There exists a constant $c_1$ such that for $\lam\geq c_1\lam_a^*$ and $\textup{Tr}(\bSig^{-1}_a)\gg s\log p$, we have with probability at least $1-2\exp(-\log p)$,
\begin{align}
&\left\|\hat{\beta}-\beta^*\right\|_1\leq C_1s\lam,~\left\|\hat{\beta}-\beta^*\right\|_2 \leq  C_2\sqrt{s}\lam,~\text{and}~~\frac{1}{\textup{Tr}(\bSig^{-1}_a)}\left\|\bX_a(\hat{\beta}-\beta^*)\right\|_2^2\leq  C_3s\lam^2\label{eq2-1}
\end{align}
for some positive constants $C_1$, $C_2$, and $C_3$. Moreover, for any $a>0$,
\[
   \lam_a^*\leq \sqrt{\frac{(\Lambda_{\max}(\Psi)/a+\sig_e^2)\log p}{\textup{Tr}(\bSig^{-1}_a)}}.
\]
\end{lemma}
\begin{remark}
\label{re3-1}
For any $a\geq 0$,
\[
  \sum_{i=1}^n \max\{m_i-q,0\}\leq \textup{Tr}(\bSig^{-1}_a)\leq N.
\]  
\end{remark}
 Lemma \ref{lem1a} provides upper bounds for the prediction error and the estimation errors in $\ell_1$-norm and $\ell_2$-norm. By setting $a$ to be a positive constant, the $\ell_2$-error of $\hat{\beta}$ is of order $\sqrt{s\log p/\textup{Tr}(\bSig^{-1}_a)}$.
 Remark \ref{re3-1} studies the magnitude of the effective sample size $\textup{Tr}(\bSig^{-1}_a)$. 
As pointed out by a reviewer,  in the case of equal group sizes and $q/m\leq c_0<1$, $\textup{Tr}(\bSig^{-1}_a)\asymp N$, i.e. the convergence rates are the same as the rates in linear models. Revoking Proposition \ref{prop1}, it shows that $\hat{\beta}$ has a faster convergence rate than $\hat{\beta}^{(lm)}$ in the regime that $q$ grows but remains relatively small to $m$. 

The results of Lemma \ref{lem1a} hold for any positive constant  $a$. Different choices of $a$ can affect the constants in the upper bound and the empirical performance of the method. To understand the optimal choice of $a$, we prove the following remark.
\begin{remark}[Effect of $a$]
\label{re3}
Suppose $\Psi=\eta^* I_q$.
For any given $n, p, q, \{m_i\}_{i=1}^n$, $a=\eta^*/\sig_e^2$ minimizes $\lam_a^*$ for $a\in(0,\infty)$. If it further holds that $\eta^*\neq 0$ and $q<\max_{i\leq n} m_i$, then $\lam^*_0> \lam^*_a$ for any $a\in(0,\lam_a^*]$.
\end{remark}
Remark \ref{re3} gives the optimal choice of $a$ for $\Psi=\eta^* I_q$. In this case, setting $a=\eta^*/\sig^2_e$ minimizes $\lam_a^*$ and hence minimizes the upper bound on the estimation errors when other parameters and constants are fixed. The optimal choice of $a$ is intuitive as it mimics the MLE procedure. 
Furthermore, when the random effects exist and $q<\max_{i\leq n} m_i$, then setting $a=0$ is strictly worse than the proposed quasi-likelihood approach with $a\in(0,\lam_a^*]$. We mention that the condition $q<\max_{i\leq n} m_i$ is sufficient but not necessary.
This remark sheds lights on the choice of $a$ in general settings as any semi-positive definite $\Psi$ can be upper and lower bounded  by diagonal matrices.
From the optimization perspective, we treat $a$ as a tuning parameter in the optimization (\ref{beta-mc}). 
In Section \ref{sec5}, we carefully examine the effect of $a$ on estimation accuracy in numerical experiments. 
 
 \subsection{Rate optimality of the proposed estimator}
 \label{sec-minimax}
In this subsection, we study the minimax optimality of proposed estimator for the fixed effects. We consider
 \begin{equation}
 \label{mini1}
   X^i_{k,.}|\bZ\sim_{i.i.d.} N\left(0,\Sig_{x}\right),~ \gam_i\sim_{i.i.d.} N(0,\Psi)~\text{and}~\eps_{i,k}\sim_{i.i.d.} N(0,\sig^{2}_e),
 \end{equation}
$k=1,\dots,m_i,~i=1,\dots,n$.  Consider the following parameter space
   \begin{align}
    \Xi(s,\bZ)&=\left\{\bm{\nu}=(\beta^*,\Psi,\sig^2_e, \Sig_{x},\bZ): \|\beta^*\|_0\leq s,~0<\sig^2_e\leq K_0,~0\preceq \Psi\preceq K_1,\right.\nonumber\\
   &\quad\quad \left. 1/K^*\leq  \Lambda_{\min}(\Sig_{x})\leq \Lambda_{\max}(\Sig_{x})\leq K^*<\infty \right\},\label{Xi}
  \end{align}
  where $K^*\geq 1$. We see that (\ref{mini1}) and (\ref{Xi}) define a special case of Conditions \ref{cond1} and \ref{cond0a}.
  We prove the minimax optimal rate of convergence in $\ell_2$-norm with respect to $\Xi(s)$.
      \begin{theorem}
 \label{thm-minimax0}\textsc{(Minimax lower bounds for estimating the fixed effects).}
 Suppose that (\ref{m1}) and (\ref{mini1}) are true. If $s\leq c\min\{\textup{Tr}(\bSig^{-1}_a)/\log p, p^{\nu}\}$ for $0<\nu<1/2$ and $c>0$, then there exists some constant $c_1>0$ such that for any fixed $a>0$,
  \begin{align*}
   &\inf_{\hat{\beta}}\sup_{\bm{\nu}\in\Xi(s,\bZ)} \P_{\bm{\nu}}\left(\|\hat{\beta}-\beta^*\|^2_2\geq  c_1\frac{s\log (p/s^2)}{\textup{Tr}(\bSig^{-1}_a)}|\bZ\right)\geq \frac{1}{4}.
 \end{align*}
 \end{theorem}
Together with (\ref{eq2-1}), this shows that $\hat{\beta}$ is minimax rate optimal in $\ell_2$-error in the parameter space $\Xi(s,\bZ)$. 
  In the proof, we use the minimax optimality of $\ell_1$-penalized MLE, which has $\bSig^{-1}_{\theta^*}$ as the weighting matrix and use Proposition \ref{prop2} to show the equivalence of MLE and proposed estimator in term of convergence rate.

\subsection{Statistical inference of the fixed effects}
\label{sec-inf}
Debiased estimators can be used for statistical inference of linear combinations of regression coefficients in high-dimensional linear models \citep{ZZ14, van14, JM14}. Under certain conditions, the debiased estimators are asymptotically normal and can be used to construct confidence interval with optimal lengths \citep{CG17}.
To make inference of $\beta^*_j$, we consider the debiased estimator proposed in (\ref{eq-db}). Let $H_j$ be the support of $(\Sig_{x|z}^{-1})_{.,j}$. 

\begin{theorem}
\label{db-clt}\textsc{(Asymptotic normality of the debiased estimator)}.
Assume Conditions \ref{cond1} and \ref{cond0a}. Let $\lam\wedge \lam_j\geq c_1\sqrt{\log p/\textup{Tr}(\bSig^{-1}_a)}$ with a large enough constant $c_1$.  For $\hat{\beta}^{(db)}_j$ defined in (\ref{eq-db}), if $(s\log p)^2\vee \log n\max_i m_i\ll \textup{Tr}(\bSig^{-1}_a)\Lambda_{\min}(\bSig_a^{-1/2}\bSig_{\theta^*}\bSig_a^{-1/2})$ and $|H_j|\log p\ll \textup{Tr}(\bSig^{-1}_a)$, then it holds that
\[
   V_j^{-1/2}(\hat{\beta}^{(db)}_j-\beta^*_j)= R_{j} + o_P(1),
\]
where $ R_{j} \xrightarrow{D}  N\left(0, 1\right)$ for 
\[
V_{j}=\frac{\hat{w}_j^{\top}\bSig^{-1/2}_a\bSig_{\theta^*}\bSig^{-1/2}_a\hat{w}_j}{\{\hat{w}_j^{\top}(\bX_a)_{.,j}\}^2}.
\]
The magnitude of $V_j$ satisfies
\begin{equation}
\label{eq-clt-re2}
V_j=\frac{(\Sig_{x|z}^{-1})_{j,j}\textup{Tr}(\bSig_a^{-1}\bSig_{\theta^*}\bSig_a^{-1})}{\textup{Tr}^2(\bSig_a^{-1})}(1+o_P(1)).
\end{equation}

\end{theorem}
  
Theorem \ref{db-clt} shows the asymptotic normality of the proposed debiased estimator under the given conditions. The convergence rate of $\hat{\beta}^{(db)}_j$ is $V^{1/2}_j$ with  magnitude  provided in (\ref{eq-clt-re2}). Remark \ref{re3} helps to understand the effect of $a$ on the inference results. As $\Sig_{x|z}$ is positive definite, $V_j$ is proportional to $(\lam^*_a)^2$ for any given $p$. Hence, using the debiased Lasso for linear models, i.e., setting $a=0$, can lead to large $V_j$ and low power in statistical inference. We verify these arguments numerically in Section \ref{sec5}.
 The sparsity of $(\Sig^{-1}_{x|z})_{.,j}$ guarantees that $\hat{\kappa}_j$ converges to its probabilistic limit so that the central limit theorem can be justified. When $\Psi$ is positive definite, the sample size condition for asymptotic normality is $(s\log p)^2\vee \log n\max_i m_i\vee |H_j|\log p\ll \textup{Tr}(\bSig^{-1}_a)$. When $\Psi$ is singular, it is sufficient to require
$(s\log p)^2\max_im_i\vee \log n\max_i m_i^2\vee |H_j|\log p\ll \textup{Tr}(\bSig^{-1}_a)$.
  
Theorem \ref{db-clt} is related to the results in some other works. 
When there is no random effect components, i.e. $\bZ=0$, the conditions and conclusions of Theorem \ref{db-clt} recover the conditions and conclusions for the debiased Lasso in linear models, say, in Theorem 2.4 of \citet{van14}. In comparison to BCG19 of \cite{Bradic17}, the limiting distribution of their test statistic under the null hypothesis does not require sparse regression coefficients but requires $|H_j|=o(\sqrt{n}/\log p/\log n)$, using our notations.  The power analysis for their test statistic requires $\max\{s,|H_j|\}=o(\sqrt{n}/\log p/\log n)$. In comparison, the sample size condition in our work (in the fixed $q$ and $m$ scenario) is $s=o(\sqrt{n}/\log p)$  and $|H_j|=o(n/\log p)$, which is weaker. More importantly, the realization of $\hat{\beta}^{(db)}_j$ does not rely on the null hypothesis and hence can be directly used to construct confidence intervals. We examine the empirical performance of these two different approaches in Section \ref{sec5}.

If one has a consistent estimator of $V_j$, the confidence intervals of the fixed effects can be constructed based on the limiting distribution of $\hat{\beta}^{(db)}_j$. However, a plug-in estimate of $V_j$ would require the knowledge of the structures of variance components and extra efforts on estimation. In the following, we show that an empirical estimator of $V_j$, $\widehat{V}_j$ defined in (\ref{eq-var-db}), is consistent under mild conditions.  Let $c_n=\log n\max_i m_i /\Lambda_{\min}(\bSig_a^{-1/2}\bSig_{\theta^*}\bSig_a^{-1/2})$.
\begin{lemma}[Convergence rate of the variance estimator]
\label{var-est}
Under the conditions of Theorem \ref{db-clt},  for $\widehat{V}_j$ defined in (\ref{eq-var-db}),
\[
  |\widehat{V}_j/V_j-1|=O_P\left(c_n^{1/2}\textup{Tr}^{-1/2}(\bSig^{-1}_a)+c_n\frac{s\log p}{\textup{Tr}(\bSig^{-1}_a)}\right).
   \]
\end{lemma}
 Lemma \ref{var-est} implies that the proposed variance estimator is consistent if $c_n=o(\textup{Tr}^{1/2}(\bSig^{-1}_a))$ and the conditions of Theorem \ref{db-clt} hold ture. The quantity $c_n$ is to account for the correlations within clusters. The proposed $\widehat{V}_j$ is robust in the sense that it does not rely on the specific structure of the variance components and is consistent under mild conditions.
Based on Theorem \ref{db-clt} and Lemma \ref{var-est}, hypothesis testing and constructing confidence intervals are both achievable. The asymptotic validness of the proposed confidence interval (\ref{ci1}) is guaranteed.

We conclude this section with a further comment on the benefits of using the quasi-likelihood. In fact, even if we have consistent estimators of the variance parameters, say $\hat{\theta}$, using proxy matrix $\bSig_a$ to compute the debiased estimator is still  favorable  over using $\bSig_{\hat{\theta}}$. First, using $\hat{\theta}$ can bring the complex dependence structure to $R_{j}$ as the correction score would also depend on the random components. This makes it difficult to justify the asymptotic normality of $R_j$. Second, $\bSig^{-1}_{\hat{\theta}}$ may not approximate $\bSig^{-1}_{\theta^*}$ well enough in the sense that the magnitude of the error in $\hat{\theta}$ can be larger than the magnitude of the bias of the debiased Lasso estimator. As a result, the sample size condition for the asymptotic normality may be impaired.

\subsection{Results for possibly correlated $\bX$ and $\bZ$}
\label{sec3-4}
In this subsection, we consider a relaxed version of Condition \ref{cond0a} that  allows for correlation between $\bX$ and $\bZ$. 
\begin{condition}
\label{cond0b}
Conditioning on $\bZ$, each row of $\bX$ is independently distributed with covariance matrix $\Sig_{x|z}$ such that $0<K_*\leq \Lambda_{\min}(\Sig_{x|z})\leq \Lambda_{\max}(\Sig_{x|z})\leq K^*<\infty$. Conditioning on $\bZ$, the conditional sub-Gaussian norms of $X^i_{k,.}$ are upper bounded by $K_0$.  Moreover, $\max_{1\leq j\leq p}\|\E[\bX_{.,j}|\bZ]\|_2^2\leq c_1\textup{Tr}(\bSig^{-1}_a)$ for some large enough $c_1$.
\end{condition}
Revoking that $\|\bX_{.,j}\|_2^2\leq CN$ and Remark \ref{re3-1}, a sufficient condition for the last statement to hold is $q/(\min_i m_i)\leq c_0<1$. 

\begin{lemma}[Fixed effects estimation with Lasso]
\label{lem1b}
Assume that Conditions \ref{cond1} and \ref{cond0b} hold true. There exist large enough constants $c_1$ and $c_2$ such that for $\lam\geq c_1\sqrt{(\sig_e^2+K_1/a)\log p/\textup{Tr}(\bSig^{-1}_a)}$ and $\textup{Tr}(\bSig^{-1}_a)\gg s\log p$, we have with probability at least $1-2\exp(-c_2\log p)$,
\begin{align}
&\left\|\hat{\beta}-\beta^*\right\|_1\leq C_1s\lam,~\left\|\hat{\beta}-\beta^*\right\|_2 \leq  C_2\sqrt{s}\lam,~\text{and}~~\frac{1}{\textup{Tr}(\bSig^{-1}_a)}\left\|\bX_a(\hat{\beta}-\beta^*)\right\|_2^2\leq  C_3s\lam^2\label{eq2-11}
\end{align}
for some large enough constants $C_1$, $C_2$, and $C_3$.
\end{lemma}
Under Condition \ref{cond0b}, for any constant $a>0$, the effective sample size for the proposed approach is still of order $\textup{Tr}(\bSig^{-1}_a)$. However, we may not have a clear understanding of the optimal choice of $a$ under current conditions but $a$ can still be chosen by cross-validation in applications.

For inference of the fixed effects, one issue caused by the correlation between  $\bX$ and $\bZ$ is that the limit of $\hat{\kappa}_j$ in (\ref{eq-kappa}) can depend on $a$ and its sparsity is hard to guarantee. If its limit is sparse indeed, then the central limit theory of Theorem \ref{db-clt} still holds. If its limit is not sparse, then one may consider the debiasing scheme for linear models with the initial estimator computed by the quasi-likelihood approach. 
We show in the Supplementary Materials (Theorem A) that our proposed debiased estimator with $a=0$ in (\ref{eq-db}) and (\ref{eq-kappa}) is robust to the correlation between $\bX$ and $\bZ$.  However, its asymptotic normality requires stronger sample size conditions when $\Psi$ is positive definite and it can have significantly wider confidence intervals and hence lower statistical power. We examine this method in Section \ref{sec5} numerically.

\section{Variance components estimation}
\label{sec4}
In this section, we consider estimating the unknown parameters of variance components. While fixed effects inference can be of main interest in many problems, estimation of variance components can provide insights into the structure of the data. As far as we know, this problem has not been studied in existence of high-dimensional fixed effects. We will demonstrate that the idea of quasi-likelihood approach can be applied to estimating the variance components in high-dimensional linear mixed-effects models. 

We parameterize $\Psi$ as follows. Let $\eta^*\in\R^d$ be the true parameter such that
\begin{equation}
\label{psi-star}
   \Psi= \Psi_{\eta^*} = \sum_{j=1}^{d}\eta^*_jG_j\in\R^{q\times q},
\end{equation}
where $G_1,\dots G_d$ are symmetric basis matrices that  are linearly independent in the sense that
\begin{equation}
\label{psi-cond1}
  \sum_{j=1}^dc_jG_j=0 ~\text{iff}~c_1=\dots=c_d=0.
\end{equation}
The dimension $d$ is allowed to grow to infinity. 
The structure of $\Psi_{\eta^*}$ (\ref{psi-star}) incorporates most commonly used models in applications, such as the random intercept model and the models used in twin or family studies  \citep{Wang11}. One should note that any symmetric $ \Psi$ can be represented via (\ref{psi-star}) with $\eta^*$ being the vector of its upper diagonal elements. Without loss of generality, we assume the basis matrices have a constant scale, i.e. $\max_{1\leq j\leq d}\|G_j\|_2 \leq C<\infty$.

\subsection{Estimating the variance components}\label{sec4-1}

A widely used approach for estimating the variance components is the Gaussian maximum likelihood method. However, this approach is highly restricted to the Gaussian assumptions. 
We consider a different approach that deals with sub-Gaussian random components. We first split the data into two folds: Let $I_1\cup I_2=[n]$, $I_1\cap I_2=\emptyset$, and $|I_1|\approx |I_2|\approx n/2$. Let $\hat{\beta}^{(2)}$ be an initial estimate of $\beta^*$ with the second half of the data $\{X^i,Z^i,y_i\}_{i\in I_2}$. We compute the residuals $\hat{r}_i=y_i-X_i\hat{\beta}^{(2)}$ for $i\in I_1$ and 
estimate $\sig^2_e$ via
\begin{align}
\label{sig-est}
   \hat{\sig}_e^2=\frac{1}{\sum_{i\in I_1}\textup{Tr}(P^{\perp}_{Z^i})}\sum_{i\in I_1}\|P_{Z^i}^{\perp}\hat{r}_i\|_2^2.
\end{align}
Next, we estimate $\eta^*$ via
\begin{align}
\label{eta-est}
  \hat{\eta}=\argmin_{\eta\in\R^d}\sum_{i\in I_1}\left\|(\Sig^i_a)^{-1/2}(\hat{r}_i\hat{r}_i^{\top}-Z^i\Psi_{\eta}(Z^i)^{\top}-\hat{\sig}^2_eI_{n_i})(\Sig^i_a)^{-1/2}\right\|_F^2,
\end{align}
where constant $K\geq K_2$ and  $\hat{\sig}^2_e$ is obtained via (\ref{sig-est}).
 
The rationale of (\ref{sig-est}) is that the observations $P_{Z^i}^{\perp}(y_i-X^i\beta^*)$ have covariance matrix $\sig^2_eP_{Z^i}^{\perp}$ which only involves the target parameter $\sig^2_e$. Replacing $\beta^*$ with its quasi-likelihood estimate gives (\ref{sig-est}). This estimator is meaningful only when $\sum_{i\in I_1}\textup{Tr}(P^{\perp}_{Z^i})>0$, i.e. $\sum_{i\in I_1}m_i\max\{0,1-q/m_i\}>0$. The rationale of (\ref{eta-est}) comes from the MLE. One can check that the derivative of the target function in (\ref{eta-est}) would be the score function with respect to $\eta$ if we replace $\bSig_a$ with the MLE estimate of $\bSig_{\theta^*}$.
Different from the MLE, we estimate $\sig^2_e$ and $\eta^*$ separately.
This is because a joint estimation of $\eta^*$ and $\sig^2_e$ may have poor performance. The reason is that, loosely speaking, the observed data involves $N$ independent observations of the random noise and $n$ independent observations of random effects. When $N\gg n$, the convergence rate for estimating $\sig^2_e$ and $\eta^*$ can have different magnitudes and a joint estimation can lead to ill-positioned Hessian matrix and non-sharp convergence rate. The sample splitting is for technical reasons and it is for proving that 
 the estimation error of $\hat{\eta}$ is independent of the error of the  fixed-effects estimation.

 Computationally, $\hat{\sig}^2_e$ in (\ref{sig-est}) is a one-step estimator and (\ref{eta-est}) involves a convex optimization, which  can be easily implemented. On the other hand, sample splitting can lead to sub-optimal finite sample performance and it is worthwhile to perform a cross-fitting step. That is, one can run another round of (\ref{sig-est}) and (\ref{eta-est}) with samples in the two folds switched and report the average of two estimates as the final estimate.

\subsection{Upper bound analysis}
In this subsection, we analyze the proposed estimator of the variance components.
Let $D_G\in\R^{d\times d}$ be such that 
\begin{equation}
\label{eq-B}
\{D_G\}_{j,k}=\textup{Tr}\left(G_jG_k\right).
\end{equation}
The matrix $D_G$ only depends on the pre-specified basis and $ \Lambda_{\min} (D_G)>0$ as $G_j$, $j=1,\dots,d$, are linearly independent.
  \begin{theorem}
  \label{thm3}\textsc{(Convergence rate of  variance components estimates)}
 Assume that Conditions \ref{cond1} and \ref{cond0a} hold and $
 \sum_{i\in I_2}\textup{Tr}((\Sig^i_a)^{-1})\gg s\log p$. Then
\[
    |\hat{\sig}^2_e-\sig^2_e|=O_P\left(\left(\sum_{i\in I_1}\max m_i\{0,1-q/m_i\}\right)^{-1/2}+\frac{s\log p}{\sum_{i\in I_2}\textup{Tr}((\Sig^i_a)^{-1})}\right).
\]
If further $n\geq c_1\log d$ for some large enough $c_1$, $\min_{1\leq i\leq n} \Lambda_{\min} (\Sig^i_z)\geq c_0/m_i>0$, and $0<c_1\leq \Lambda_{\min} (D_G)\leq \Lambda_{\max}(D_G)\leq c_2<\infty$, then 
\begin{align*}
\|\hat{\eta}-\eta^*\|_2=O_P\left(\sqrt{\frac{d\log d}{n}}\right).
\end{align*}
  \end{theorem}  
  The convergence rate of $\hat{\sig}^2_e$ depends on the effective sample size in $I_1$ as well as the estimation error of $\hat{\beta}^{(2)}$. In comparison to the rate of variance estimation in linear models \citep{Verzelen12}, the current result replaces the total sample size with the effective sample size.
On the other hand, $\hat{\eta}$ has the typical parametric rate when there are $d$ unknown parameters and $n$ independent observations of random effects. The estimation error of $\hat{\eta}$ is independent of   the error of the fixed effects estimation.

In terms of conditions,
$D_G$ depends on pre-specified basis matrices and it eigenvalues are positive and bounded in many cases. Consider the class of basis matrices where $G_{j,k}\in\R^{q\times q}$ such that $(G_{j,k})_{l,q}=(G_{j,k})_{q,l}=1$ if $l=j,q=k$ and $(G_{j,k})_{l,q}=0$ otherwise. For any $1\leq d\leq q(q+1)/2$, it is easy to check that in this case $\Lambda_{\min}(D_G)=\Lambda_{\max}(D_G)=1$. For independent sub-Gaussian $Z^i_{k,.}$, $\min_{1\leq i\leq n} \Lambda_{\min} (\Sig^i_z)\geq c_0$ with high probability if $q\log n\ll \min_{1\leq i\leq m} m_i$. To summarize, the estimators proposed in Section \ref{sec4-1} are mainly for the scenario where $q\leq c_0 \min_{1\leq i\leq n}m_i$ for sufficiently small $c_0$.

 \subsection{Rate optimality of estimating variance components}
 \label{sec-minimax2}

 Now we turn to study the minimax lower bound for estimating the variance parameters. We consider the random components satisfying (\ref{mini1}) and parameter space $\Xi(s,\bZ)$ (\ref{Xi}).
  \begin{theorem}
 \label{thm-minimax1}\textsc{(Minimax lower bounds for estimation of variance components.)}
 Suppose that (\ref{m1}) and (\ref{mini1}) are true. If $s\leq c\min\{\textup{Tr}(\bSig^{-1}_a)/\log p, p^{\nu}\}$ for $0<\nu<1/2$ and $c>0$ for some $c_0>0$, then there exists some constants $c_1-c_3>0$ such that
   \begin{align*}
  &\inf_{\hat{\sig}^2_e}\sup_{\bm{\nu}\in\Xi(s,\bZ)} \P_{\bm{\nu}}\left(|\hat{\sig}^2_e-\sig^{2}_e|\geq  c_1\textup{Tr}^{-1/2}(\bSig^{-1}_a)+c_2\frac{s\log (p/s^2)}{\textup{Tr}(\bSig^{-1}_a)}|\bZ\right)\geq \frac{1}{4}
  \end{align*}
  If further $\Lambda_{\max}(D_G)\leq C<\infty$,
  \begin{align*}
 &\inf_{\hat{\eta}}\sup_{\bm{\nu}\in\Xi(s,\bZ)} \P_{\bm{\nu}}\left(\|\hat{\eta}-\eta^*\|_2\geq  c_3n^{-1/2}|\bZ\right)\geq \frac{1}{4}.
 \end{align*}
 \end{theorem}
  Theorem \ref{thm-minimax1} and Theorem \ref{thm3} together imply that $\hat{\sig}^2_e$ is rate optimal in $\Xi(s,\bZ)$ under the conditions of Theorem \ref{thm-minimax1} if when $\textup{Tr}(\bSig^{-1}_a)\asymp \sum_{i=1}^n m_i\max\{0,1-q/m_i\}$. As explained after Lemma \ref{lem1a}, in the case where group sizes are equal and $q/m\leq c_0<1$, $\sum_{i=1}^n \max\{0,m_i-q\}\asymp \textup{Tr}(\bSig^{-1}_a)\asymp N$.
  Moreover, $\hat{\eta}$ is rate optimal when $d$ is finite. When $d$ grows, regularized estimators of $\eta^*$ can have smaller estimation error than $\hat{\eta}$, similar to  the famous Stein's phenomenon.

\section{Simulation results}
\label{sec5}
In this section, we present simulation results to evaluate the empirical performance of the proposed methods and compare it with some related methods.  We examine the effect of $a$ on estimation and inference of the fixed effects. 

We generate data as follows.   We set $N=144$ and $p=300$. Each row of $(\bX,\bZ)$ are i.i.d. generated from a normal distribution with mean zero and covariance such that $\Sig_x=I_p$, $\Sig_z=I_q$, and $(\Sig_{x,z})_{k,j}=\rho^{j}$ for $1\leq j,k\leq q$ and $(\Sig_{x,z})_{k,.}=0$ for $k>q$.
That is, the correlation between $\bX_j$  and $\bZ$ is nonzero if $j\leq p$ and is $0$ if $j> q$. The random noises are \textit{i.i.d.} generated via $\eps_{i}\sim N(0, 0.25I_{m_i})$ and the random effects are \textit{i.i.d.} generated via $\gam_i\sim N(0,\Psi)$. We consider $q\in\{2,8,14\}$. The matrix $\Psi$ will be specified later. The responses $\by$ are generated via model (\ref{m1}) with  $s=5$ and $\beta_{1:5}=(1,0.5,0.2,0.1,0.05)^{\top}$ and  equal cluster sizes, i.e. $m_1=\dots=m_n=m$. 
 Each setting is replicated with 300 independent Monte Carlo simulations.

\subsection{Statistical inference for fixed effects}
\label{sec5-1}
We first examine the empirical performance of the proposed confidence intervals (\ref{ci1}) and hypothesis testing based on $\hat{\beta}^{(db)}_j$.
We consider two covariance matrices of random effects, a ``positive definite $\Psi$'' where $\Psi_{j,k}=0.56^{|j-k|}$ for $1\leq j,k\leq q$, and  a ``singular $\Psi$''  with a diagonal $\Psi$ where $\Psi_{j,j}=0.56$ for $1\leq j\leq q/2$ and $\Psi_{j,j}=0$ otherwise.  
For the proposed method, we first choose $a$ by cross-validation using the error criteria $\|\by-\bX\hat{\beta}(a)\|_2^2$, where $\hat{\beta}(a)$ the the proposed estimate associated with a specific $a$. The tuning parameter $\lam$ is chosen as 
$\hat{\sig}^{(init)}\sqrt{2\log p/ N}$, where $\hat{\sig}^{(init)}$ is computed via the scaled-Lasso  \citep{scaled-lasso} with observations $\{\bX_a,\by_a\}$. For computing $\hat{\beta}^{(db)}_j$, the tuning parameters $\lam_j$ are set to be $\hat{\sig}_x\sqrt{2\log p/ N}$, where $\hat{\sig}_x$ is computed via the scaled-Lasso with observations $\{(\bX_a)_{.,-j},(\bX_a)_{.,j}\}$. The tuning parameters for BCG19 are chosen as in Section 5 of \citet{Bradic17}.

We see from Table \ref{table-inf} that the coverage probabilities of the proposed confidence intervals  are close to the nominal level in most scenarios. It shows that the proposed method is robust to large $m$ and $q$ and singular $\Psi$. We see that the confidence intervals have shorter lengths when $m$ increases. This is because when $q$ is fixed and $m$ grows, the effective sample size $\textup{Tr}(\bSig_a^{-1})$ increases and the proposed estimators have smaller estimation errors. See Table 1 in the Supplementary Materials for details. When $q$ grows and $m$ is fixed, the effective sample size $\textup{Tr}(\bSig_a^{-1})$ is smaller and the proposed estimators have larger estimation errors. The results for $\rho=0.2$ are reported in the Table 2 of the Supplementary Materials.

\begin{table}[!htbp]
	\caption{95\%-confidence intervals given by the proposed approach with positive definite $\Psi$ and singular $\Psi$ when $\rho=0$.``cov(0.5)'' and ``cov(0)'' denote the coverage probabilities for $\beta_j=0.5$ and $\beta_j=0$, respectively. ``sd(0.5)'' and ``sd(0)''  denote the standard deviations for $\beta_j=0.5$ and $\beta_j=0$, respectively.\label{table-inf} }
		\renewcommand{\arraystretch}{0.9}
	\begin{center}
	{\small
	\begin{tabular}{|c|c|cccc|cccc|}
		\hline
		\multirow{2}{*}{$q$}&\multirow{2}{*}{$m$}&\multicolumn{4}{c|}{Positive definite $\Psi$} &  \multicolumn{4}{c|}{Singular $\Psi$} \\
		\cline{3-10}
		& & cov(0.5) & cov(0) & sd(0.5)&sd(0)  & cov(0.5) & cov(0) & sd(0.5)&sd(0)  \\
		\hline
		\multirow{3}{*}{2} &4 & 0.940& 0.957 & 0.068&0.062 &0.953& 0.943 & 0.062 & 0.056 \\
		
		& 8 & 0.943&  0.967& 0.063& 0.053  &0.938&0.981 & 0.058 & 0.049 \\
		
		&12 &0.943&  0.967& 0.061 & 0.049& 0.967&0.948  & 0.059 & 0.047 \\
		\hline
		\multirow{3}{*}{8}&4 & 0.943 & 0.960 & 0.195&0.177 & 0.957 &0.943  & 0.128& 0.111\\
		
		&8 & 0.960 &  0.940& 0.148& 0.123  & 0.933&0.919  & 0.105 &  0.088 \\
		
		&12 & 0.943 &  0.973&  0.106 & 0.083&0.957&0.976&0.085&0.066 \\
		\hline
		\multirow{3}{*}{14}&4 & 0.937 & 0.953 & 0.276&0.264 & 0.976 &0.948 & 0.173 & 0.153\\
		
		&8 & 0.937 &  0.947& 0.243& 0.217 & 0.924 &0.929  & 0.158 &  0.132 \\
		
		&12 & 0.933& 0.950&  0.202 & 0.166&0.981 &0.924&0.148 &0.112 \\
		\hline
	\end{tabular}
}
\end{center}	
\end{table}

In Table \ref{table-inf2}, we report the type-I  error and power of our proposed method and those of BCG19. The computational time for our proposal is around 8s per experiment and that for BCG19 is around 20s per experiment. Ideally, the rejection rate for the true null should be close to 5\% and the rejection rates for $\beta^*_j\in\{1,0.5,0.2\}$ should be larger than $5\%$. We see that both our proposal and BCG19 are effective at controlling the type-I error. However, BCG19  is less powerful than our proposal when $q$ is large and $\Psi$ is positive definite. When $\Psi$ is singular, two methods have comparable performance in most scenarios. 

 \begin{table}[!htbp]
	\caption{The rejection rate for testing $H_0: \beta^*_j=0$ at 95\% level for $\beta^*_j\in\{1,0.5,0.2,0\}$ with positive definite $\Psi$ (p.d.) and singular $\Psi$ when $\rho=0$.\label{table-inf2} }
	\begin{center}
	\renewcommand{\arraystretch}{0.9}
	{\small
	\begin{tabular}{|c|c|c|cccc|cccc|}
		\hline
  \multirow{2}{*}{$\Psi$}& \multirow{2}{*}{$q$}&\multirow{2}{*}{$m$}& \multicolumn{4}{c|}{Proposed} & \multicolumn{4}{c|}{BCG19} \\
		\cline{4-11}
		&& &1 & 0.5 & 0.2& 0 & 1  & 0.5 & 0.2&  0 \\
		\hline
\multirow{9}{*}{p.d.} &\multirow{3}{*}{2}&4 & 1&1& 0.793&0.043 & 1 &  1 & 0.627 & 0.043\\
		
	&	& 8 &1 &  1 & 0.880 & 0.033 &1&1  &0.850 & 0.040 \\
		
	&	&12 &1&  1& 0.940&  0.033 & 1 &1  & 0.936    & 0.040\\
	\cline{2-11}
	&	\multirow{3}{*}{8}&4&0.997&  0.713&  0.163 & 0.040 & 0.987 &0.593 & 0.150 & 0.067\\\
		
	&	&8     &1  & 0.923&  0.243 & 0.060 & 0.987&0.747  & 0.173&  0.060 \\
		
	&	&12   &1  & 1&  0.423 & 0.027 & 1&0.840  & 0.207      &  0.030 \\
	\cline{2-11}
	&  \multirow{3}{*}{14}&4 &0.943&  0.437 &  0.117& 0.047 & 0.867 &0.327 & 0.123 & 0.037\\
		
	&	&8 & 0.967 & 0.487&  0.113 & 0.053 & 0.927&0.397 & 0.107 &  0.057 \\
		
	&	&12& 0.993& 0.610 &  0.157 & 0.050 & 0.930 & 0.477& 0.150 & 0.060\\
		\hline
	\multirow{9}{*}{singular} &\multirow{3}{*}{2}&4 & 1& 1 & 0.895&0.057 &1 &1 & 0.900& 0.029\\
		
	&	& 8 &1 &  1 & 0.952 & 0.020 &1&1  &0.943 & 0.024 \\
		
	&	&12 &1&  1& 0.914&  0.052& 1 &1& 0.927 & 0.057\\
			\cline{2-11}
	&	\multirow{3}{*}{8}&4 &1 &0.995 &0.362 &0.057 &1 &0.976 &0.371 &0.052  \\
		
	&	&8 & 1 &0.986 &0.438 &0.071&1 &0.986 &0.438&0.062 \\
		
	&	&12 & 1 &  1 &  0.638 &0.024 &1 &1 &0.567 &0.047\\
		\cline{2-11}
	& \multirow{3}{*}{14}&4 &1&1& 0.638 &0.024 &1&1&0.567 &0.048\\
		
	&	&8 &1&0.895 &0.228 &0.043 &1 &0.890 &0.233 &0.052\\
		
	&	&12 &1&  0.895 &  0.229& 0.067 &1 &0.890&0.233 &0.052 \\
		\hline
	\end{tabular}
}
\end{center}	
\end{table}

Table \ref{table-a} demonstrates the effect of $a$ on estimation and inference of the fixed effects. The results for singular $\Psi$ are reported in the Supplementary Materials. We see that choosing $a=0$ can lead to large estimation errors and significantly wider confidence intervals. This implies that the Lasso for linear models is less accurate than our proposed methods with $a>0$.  In all the scenarios of $(q,m)$, we see that the estimation error first decreases as $a$ increases and then increases as $a$ increases. This phenomenon agrees with Remark \ref{re3}. For the inference results, the proposed confidence interval has the desired coverage probabilities as long as $a$ is not too large. We see that setting $a=0$ has coverage probabilities close to the nominal level but the confidence intervals are significantly wider than setting $a>0$. This implies that using the linear debiased Lasso can lead to low power in hypothesis testing for mixed-effects models.

\begin{table}[!htbp]
	\caption{Effect of different $a$ on sum of squared error (SSE) for estimating $\beta^*$ and on the accuracy of confidence intervals with positive definite $\Psi$ and $\rho=0$. ``cov(0.5)'' and ``cov(0)'' denote the coverage probabilities for $\beta_j=0.5$ and $\beta_j=0$, respectively. ``sd(0.5)'' and ``sd(0)''  denote the standard deviations for $\beta_j=0.5$ and $\beta_j=0$, respectively.\label{table-a} }
		\begin{center}
		\renewcommand{\arraystretch}{0.9}
		{\small
	\begin{tabular}{|c|c|cccccc|}
		\hline
$(q,m)$	& $a$	&SSE & $\textup{Tr}(\bSig^{-1}_a)$ &cov(0.5)  & cov(0) & sd(0.5)  & sd(0)   \\
	\hline
\multirow{7}{*}{(2,4)} &    0  & 0.321 & 144.0 & 0.948 & 0.967 & 0.138 & 0.122\\ 
 &    2 & 0.134 & 87.3 & 0.962 & 0.962 & 0.074 & 0.065\\ 
   &    4 & 0.113 & 81.4 & 0.933 & 0.952 & 0.070 & 0.062\\ 
    &    8 & 0.111 & 77.7 & 0.933 & 0.957 & 0.068 & 0.062 \\
 &   16 & 0.103 & 75.2 & 0.900 & 0.967 & 0.065 & 0.060 \\ 
    &   32 & 0.105 & 73.8 & 0.890 & 0.933 & 0.065 & 0.060\\ 
  \hline
\multirow{7}{*}{(8,8)} &  0  & 0.753 & 144.0 & 0.943 & 0.943 & 0.261 & 0.235\\ 
 &    2 & 0.338 & 34.5 & 0.952 & 0.962 & 0.150 & 0.122\\ 
 &    4 & 0.342 & 26.2 & 0.948 & 0.933 & 0.148 & 0.121 \\ 
 &    8 & 0.349 & 19.6 & 0.943 & 0.967 & 0.144 & 0.119\\ 
 &   16  & 0.350 & 14.8 & 0.910 & 0.962 & 0.143 & 0.116\\ 
&   32 & 0.402 & 10.9 & 0.895 & 0.967 & 0.142 & 0.119\\
  \hline
\multirow{7}{*}{(14,12)} &    0 & 0.961 & 144.0 & 0.948 & 0.967 & 0.344 & 0.316 \\ 
 &    2  & 0.531 & 19.9 & 0.981 & 0.948 & 0.211 & 0.167\\ 
 &    4 & 0.501 & 12.9 & 0.938 & 0.967 & 0.199 & 0.162\\ 
 &    8 & 0.515 & 8.1 & 0.938 & 0.971 & 0.200 & 0.167 \\ 
 &   16 & 0.504 & 5.0 & 0.948 & 0.957 & 0.200 & 0.161\\ 
&   32 & 0.551 & 2.9 & 0.905 & 0.967 & 0.190 & 0.155 \\ 
\hline
	\end{tabular}
	}
\end{center}	
\end{table}


\subsection{Estimating variance components}
In this subsection, we consider estimating variance components with the proposed method. The true fixed effects and data generation steps are the same as in Section \ref{sec5-1}. We use the whole data to estimate $\sig^2_e$ and $\eta^*$. We set $\sig^2_e=0.25$. We first consider diagonal $\Psi$ with $d=2$. 
The basis matrices are set to be
$$G_1=\begin{pmatrix} I_{q/2} & 0 \\
0 & 0
\end{pmatrix}~\text{and}~ G_2=\begin{pmatrix} 0 & 0\\
0 & I_{q/2}
\end{pmatrix}.$$
For diagonal $\Psi$, $\eta^*=(0.56,0.56)^{\top}$. For singular $\Psi$, $\eta^*=(0.56,0)^{\top}$. 
Table \ref{table-var1} shows the mean absolute errors of $\sig^2_e$ (mae.$\sig^2_e$), $\eta^*_1$ (mae.$\eta_1$), and $\eta^*_2$ (mae.$\eta_2$). A scenario with relatively large $d$ is reported in the Supplementary Materials (Table 4).
\begin{table}
\caption{Estimation of the variance components with the proposed method for positive definite and singular $\Psi$ when $\rho=0$. \label{table-var1} }
\begin{center}
\renewcommand{\arraystretch}{0.9}
\small{
\begin{tabular}{|c|c|ccc|ccc|}
\hline
\multirow{2}{*}{$m$}&\multirow{2}{*}{$q$} & \multicolumn{3}{c|}{Positive definite $\Psi$}  & \multicolumn{3}{c|}{Singular $\Psi$}\\
\cline{3-8}
& &mae.$\sig^2_e$ & mae.$\eta_1$ & mae.$\eta_2$& mae.$\sig^2_e$ & mae.$\eta_1$ & mae.$\eta_2$\\
\hline
4&2&  0.115 & 0.206 &0.207& 0.076 & 0.150 & 0.050\\
\cline{1-2}
\multirow{2}{*}{8} & 2  & 0.091 & 0.212 &0.249 & 0.070& 0.171& 0.020 \\
\cline{2-2}
&4  &0.122  & 0.164 & 0.166 &0.071 & 0.136 & 0.027 \\
\hline
\multirow{2}{*}{12}&2  & 0.087 & 0.268 & 0.245& 0.076& 0.197 & 0.015 \\
\cline{2-2}
&6   & 0.126 & 0.163 & 0.160 & 0.078 & 0.116 &0.019 \\
\hline
\end{tabular}
}
\end{center}
\end{table}

\section{Application to a genome-wide association study in a mouse population}
\label{sec6}
\begin{figure}
	\begin{center}
	\includegraphics[height=0.25\textwidth, width=0.5\textwidth]{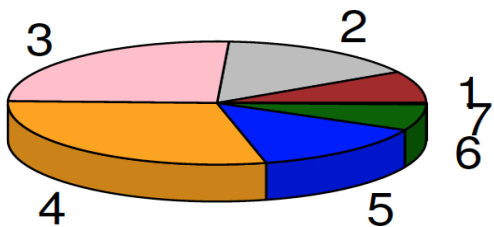}
	\end{center}
	\caption{ Cage density of the stock mice population.  The average density of cages with at least two individuals is 3.70. }
	\label{fig1}
\end{figure}

We apply the proposed method to estimate the effects of genetic variants on the body mass index (BMI) in a heterogenous stock mice population generated  by the Welcome Trust Centre for Human Genetics \url{http://gscan.well.ox.ac.uk}. The data is available in R package ``BGLR'' \citep{BGLR}.  The dataset consists of 1,814 mice, each genotyped over 10,346 polymorphic markers (SNPs) and has been used for genome-wide genetic association studies of multiple traits \citep{Mott06, Flint06}. This  mice population consists of 8 liters and were housed in 523 different cages, each including a different number of mice. The distribution of cage density is in Figure \ref{fig1}.  We are interested in identifying the genetic variants that are associated with the BMI phenotype. The measurements of BMI are transformed as described in \citet{Flint06} so that the data distribution is close to normal. In many mice experiments,  cages often contribute significant environmental effects to the phenotypes such as BMI and mice in the same cage tend to be correlated in their phenotype measurements. It is therefore important to account for such cage effect in genetic association studies and the linear mixed-effects model can be employed.

 In the current analysis, we incorporate the effect of cages as a single random effect and consider the following model 
$$Y_{i,k}=\beta_0+ \sum_{j=1}^{10346} \beta_jX^i_{j,k} + \tau_1 age_{i,k} + \tau_2  gender_{i,k} + \gam_i+ \epsilon_{i,k},$$
where $Y_{i,k}$ is the BMI of the $k$th mouse in the $i$th cage, $X^i_{j,k}$ is the numerical genotype at the genetic variant $j$ for the $i$th mouse in cage $k$,  $\beta_0$ and  $\beta_j$ are the regression coefficients corresponding to the intercept and genetic variants, $\tau_1$ and $\tau_2$ are the regression coefficients for age and gender, $\gam_i$ is the cage-specific random effect for the $i$-th cage.  
 For cages with only one individual, we only fit the fixed effects.

The fixed effects are estimated via a weighted Lasso. To mitigate the relatively high correlation among the design, we first compute ridge regression estimates of the fixed effects, say $\hat{\beta}^{(rr)}$,  with tuning parameter chosen by cross-validation and use normalized $\{1/|\hat{\beta}_j^{(rr)}|\}_{j=1}^p$ (sum up equal to $p$) as the weights for the Lasso estimates. 
The regression coefficient $\hat{\beta}$ is obtained by fitting (\ref{beta-mc}) with tuning parameter $\lam=0.655\times \sqrt{2\log p/N}$, where $0.655$ is the noise level estimated by the scaled Lasso. In terms of statistical inference, we compute the debiased Lasso estimates of the fixed effect via (\ref{eq-db}) and their variances according to (\ref{eq-var-db}). According to cross-validation, we set $a=2$.

\subsection{Identification of BMI associated genetic variants}
\begin{figure}
	\begin{center}
		\includegraphics[height=0.35\textwidth, width=0.48\textwidth]{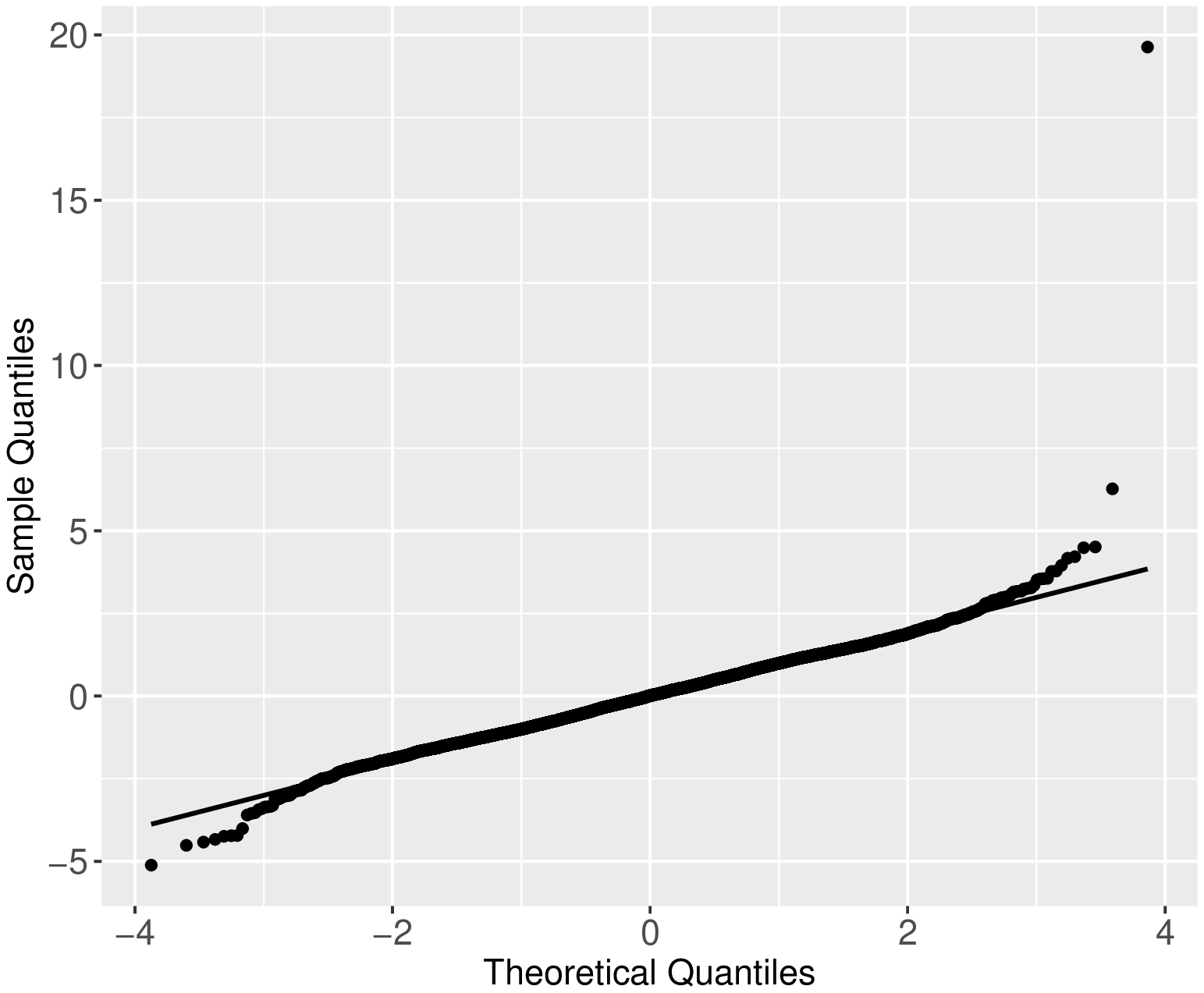}
		\includegraphics[height=0.35\textwidth, width=0.48\textwidth]{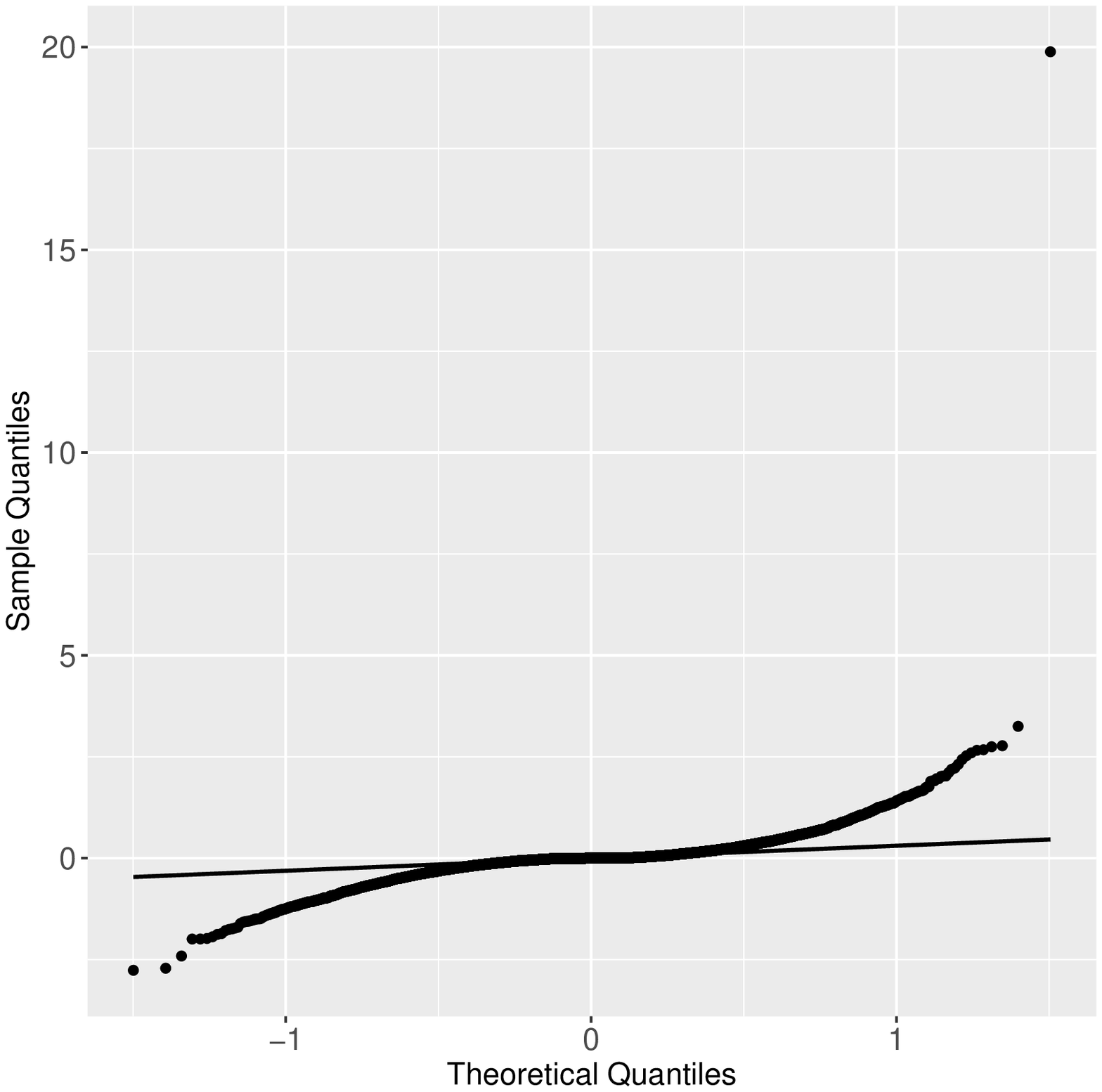}
	\end{center}
	\caption{The normal QQ-plots of the $z$-scores of  the debiased Lasso estimators with proposed approach (left) and the debiased Lasso estimates with  $a=0$ (right) for 10348 fixed effects. The straight reference line passes the first and third quantiles of the $z$-scores. \label{fig2} }
\end{figure}

 We  control the false discovery rate (FDR) at 5\% using  the procedure proposed in \citet{Xia18}. Our method identifies 14 covariates with $p$-value threshold $6.7\times 10^{-5}$. The QQ plot of the $z$-scores of all the covariates is  given in the left panel of Figure \ref{fig2}. It shows some deviation from standard normal density at both tails, indicating that some variants can be associated with BMI.  Some of the genetic variants identified are in or near the genes known to be associated with body growth, body size, metabolism or obesity.  For example,  SNP \texttt{rs13478535} is a variant in  Auts2 gene, which has been shown to be related to with either low birth weight or small stature mice  \citep{Gao14}. 
SNP \texttt{rs13481413} is one of the genetic variants in gene  Immp2l,  which is associated with food intake and body weight \citep{ Immp2l}. cAMP response element binding protein (Crebbp) has been postulated to play an important role downstream of the melanocortin-4 receptor and may affect other pathways that are implicated in the regulation of body weight \citep{crebbp}. 

We also consider applying the proposed procedure with $a=0$. This is equivalent to applying the Lasso  to fit the linear model to without considering the random cage effects.  The tuning parameters are chosen in the same way as above. Only \texttt{gender} is selected as nonzero at FDR level 0.05. This is possibly due to the model misspecification and larger variances of the debiased Lasso estimators. The QQ-plot of the $z$-scores based on the de-biased Lasso estimation of the linear model (right panel of Figure \ref{fig2}) shows that the $z$-scores clearly deviate from the standard normal distribution.   These results indicate that the proposed estimation and inference methods for the linear mixed-effects model  indeed provide an effective way of identifying important genetic variants associated with BMI in mice. 

  \begin{table}[ht]
	\caption{Selected covariates at FDR level 0.05. 13  SNPs and 
		\texttt{gender} are selected at FDR $\leq 0.05$ and their $z$-scores are reported. The genes where the SNPs are located are presented when they are available. \label{table-mice}} 
	\centering
	\begin{tabular}{|c@{\hskip 0.2cm}c@{\hskip 0.2cm}c|c@{\hskip 0.2cm}c@{\hskip 0.2cm}c|}
		\hline
		SNP & Gene&$z$-score &  SNP &Gene& $z$-score \\
		\hline
 
		\texttt{rs13476390}&  &-5.11 &\texttt{rs13482464} & &-4.01\\
		\hline
		\texttt{rs13478535} &Auts2 & -4.42&  \texttt{rs4152477} &Crebbp&4.17\\
		\hline
		\texttt{rs3023058} &Srrm3 & 4.22 & \texttt{rs6251709} & Csnka2ip&4.49\\
		\hline
		\texttt{rs13480072} && -4.52 & \texttt{rs6185805} &Mtcl1& -4.22\\
		\hline
		\texttt{rs13481413} & Immp2l&-4.22 &\texttt{gnf18.028.738} & & -4.34\\
		\hline
		\texttt{rs13481961} &&6.27&\texttt{gnfX.070.167} & &-4.24\\
		\hline
		\texttt{rs4139535} && 4.51&\texttt{gender} & &19.63\\
		\hline
	\end{tabular}
	
\end{table}

\subsection{Evaluation of cage effect}
For estimating the variance components, we only use the clusters with at least two observations. The estimated variance of the random effects is 0.202 and the estimated variance of the noise is 0.209. We compute the standard error of the estimated variance of the random effects assuming that the random components are normally distributed. The estimated standard deviation is 0.018, which indicates a strong cage effect. 




\section{Discussion} 
\label{sec-diss}
The present paper considers estimation and inference of unknown parameters in a high-dimensional linear mixed-effects model. Optimal rate of convergence for estimation was established and rate-optimal estimators were developed. The proposed methods have general applicability in modeling repeated measures and longitudinal data, especially when the cluster sizes are large or heterogeneous. The desirable properties of the proposed estimators are mainly due to the proper approximations of the unknown oracle weighting matrix $\bSig_{\theta^*}$. Our proposed estimation procedure is computationally efficient and does not require strong distributional assumptions on the random effects and error distributions.  

The proposed methods have important applications in large-scale genetic association studies in humans, including both family-based studies where the kinship coefficients can be used to specify the random effects and population cohort studies where the random effects can be used to adjust for population stratification \citep{Yang2014}.  Instead of considering one genetic variant at a time as in typical mixed-effects models in genetic association studies \citep{Yang2014}, our model considers all the variants jointly. We expect gain in power in detecting phenotype-associated genetic variants  by allowing for  flexible random effects and by considering all genetic variants jointly using  high-dimensional mixed-effects models studied in this paper.

\section*{FUNDING}
This research was supported by NIH grants R01GM123056 and R01GM129781. Tony Cai's research was also supported in part by NSF grants DMS-1712735 and DMS-2015259.

\section*{SUPPLEMENTARY MATERIALS}
In the online Supplemental Materials, we provide proofs of all the theorems and lemmas and more numerical studies.

\bibliography{HD-mix}
\bibliographystyle{chicago}

\end{document}